\documentclass[prl,twocolumn,showpacs,amsmath]{revtex4}

\usepackage{graphicx}
\usepackage{dcolumn}

\newcommand{\ee}      {\mbox{e-e}}
\newcommand{\Hee}     {\ensuremath{H_\text{e-e}}}
\newcommand{\TB}      {TB}

\begin{document}


\title{
Electron-electron interaction effects on optical excitations in 
semiconducting single-walled carbon nanotubes}

\author{Hongbo Zhao}
\author{Sumit Mazumdar}
\affiliation{Department of Physics, University of Arizona, Tucson, AZ 85721}
\date{August 16, 2004}

\begin{abstract}
  We report correlated-electron calculations of optically excited
  states in ten semiconducting single-walled carbon nanotubes with
  a wide range of diameters. Optical excitation occurs to excitons
  whose binding energies decrease with the increasing nanotube
  diameter, and are smaller than the binding energy of an isolated
  strand of poly-(paraphenylene vinylene).  The ratio of the energy of
  the second optical exciton polarized along the nanotube axis to that
  of the lowest exciton is smaller than the value predicted within
  single-particle theory.  The experimentally observed weak
  photo\-luminescence is an intrinsic feature of semiconducting
  nanotubes, and is consequence of dipole-forbidden excitons occurring
  below the optical exciton.
\end{abstract}

\pacs{78.67.Ch, 78.55.-m, 71.35.-y}


\maketitle

Recent experiments in semiconducting single-walled carbon nanotubes
(SWCNTs) have indicated the strong role of electron-electron (\ee)
interactions
\cite{Ichida-02-prb,Liu-02-prb,Bachilo-02-science,OConnell-02-science,
  Korovyanko-04-prl}, ignored in early one-electron theories
\cite{single-particle}.
Several observations have attracted particular attention.
First, 
optical gaps in SWCNTs are considerably greater
\cite{Ichida-02-prb,Liu-02-prb,Bachilo-02-science} than those
predicted from the tight-binding (\TB) model \cite{single-particle}.
Second, 
the ratio of the threshold energy corresponding to the second optical
transition polarized along the SWCNT axis to that of the first such
transition is less than the value 2
\cite{Liu-02-prb,Bachilo-02-science,OConnell-02-science} predicted
within the \TB\ model for wide SWCNTs \cite{single-particle}.
It has been claimed that this ``ratio problem'' is a signature of \ee\ 
interactions \cite{KaneMele-03-prl}.  
Third, 
ultrafast pump-probe spectroscopy has revealed {\it structured}
photo\-induced absorptions (PA) and correlations of PA with photo\-induced
bleaching (PB), that indicate that photo\-excitations in SWCNTs are
excitons \cite{Korovyanko-04-prl}.  
These observations have led to theoretical studies of SWCNTs that go
beyond one-electron models
\cite{KaneMele-03-prl,Ando-97-jpsj,Lin-00-prb,Spataru-04-prl,Chang04}.
Different calculations have, however, focused on different approaches
and often on specific SWCNTs, and while a consensus is emerging that
optical absorptions in semiconducting SWCNTs are due to excitons, 
complete physical understanding of the generic effects of \ee\ 
interactions is still missing.

In the present Letter, we investigate SWCNTs within the semiempirical
Pariser-Parr-Pople (PPP) $\pi$-electron 
Hamiltonian
\cite{PPP} that has been used extensively to discuss 
$\pi$-conjugated polymers
\cite{Chandross,Ramasesha,Rice}, the {\it other} class of
quasi-one-dimensional $\pi$-conjugated systems that exhibit strong
excitonic features \cite{Leng,Weiser}.  The advantages of the
semiempirical approach are, (i) immediate connection to the rich
physics of $\pi$-conjugated polymers can be made, and (ii) the
dominant effects of \ee\ interactions in SWCNTs can be understood
physically.
Admittedly, $\pi$-electron only theory will miss the curvature effects
of the narrowest tubes, but our emphasis is on generic
conclusions valid also for the widest tubes.

Our work has multiple conclusions.
First, we show theoretically that the observed low quantum efficiency (QE) 
of the photo\-luminescence (PL) of SWCNTs 
\cite{OConnell-02-science,Lebedkin03,Wang04,Sheng04}  
is very likely a consequence of 
the occurrence of optically forbidden exciton states below the optically 
allowed exciton.
Second, while transverse photo\-excitations are not expected to be
strongly visible in optical measurements \cite{AjikiAndo94}, the
energetics of these states are nevertheless of interest. We show that
while within the \TB\ theory the transverse
photo\-excitations occur exactly halfway between the two lowest
longitudinally polarized absorptions, 
they are shifted to considerably
above the central region. Importantly, both these results could have
been anticipated from previous work on
poly-paraphenylenevinylene (PPV) \cite{Chandross,Rice,Chandross1}.
Third, we show that the ``ratio problem'' can be understood at
the level of mean-field theory of \ee\ interactions, and no
sophisticated many-body explanation \cite{KaneMele-03-prl} is
necessary. 
Finally, we present descriptions of the optically allowed excitons in
ten different SWCNTs with diameters \mbox{5.6--13.5 \AA} to arrive at 
generic conclusions about the underlying excitonic
electronic structures.


We consider the PPP model Hamiltonian \cite{PPP}
\begin{subequations}
\label{eq:Hamiltonian}
\begin{equation}
H = H_\text{1e} + \Hee ,                            \label{eq:H}
\end{equation}
where $H_\text{1e}$ is the one-electron H\"uckel Hamiltonian and 
\Hee\ is the \ee\ interaction,  
\begin{align}
  H_\text{1e} & = - t \sum_{\langle ij \rangle, \sigma}
                      c_{i,\sigma}^\dag c_{j,\sigma} + \text{H.c.} ,  
                                                    \label{eq:H1e} \\
  \Hee & = U \sum_i n_{i,\uparrow} n_{i,\downarrow} + 
                      \frac{1}{2}\sum_{i,j} V_{ij} (n_i-1)(n_j-1)
                                                    \label{eq:Hee} .
\end{align}
\end{subequations}
%
%
Here $c_{i,\sigma}^\dag$ creates a $\pi$-electron of spin $\sigma$ on
carbon (C) atom~$i$, {\scriptsize $\langle \cdot\cdot \rangle$}
denotes nearest neighbors, $n_i=\sum_\sigma c_{i,\sigma}^\dag
c_{i,\sigma}$ is the total number of $\pi$~electrons on site $i$. The
parameters $t$, $U$ and $V_{ij}$ are the nearest-neighbor hopping
integral, and the on-site and inter-site Coulomb interactions,
respectively. We have chosen the standard value of 2.4 eV for $t$
\cite{Chandross,Ramasesha}.  Our parametrization of the long-range
$V_{ij}$ is similar to the standard Ohno parametrization
\cite{Ohno-64}
\begin{equation}
 V_{ij} = \frac{U}{\kappa \sqrt{1+0.6117 R_{ij}^2}} ,
\label{eq:Vij}
\end{equation}
where $R_{ij}$ is the distance between C atoms $i$ and $j$ in \AA, and
$\kappa$ is a screening parameter ($\kappa$ = 1 within  
Ohno parameterization) 
\cite{Chandross1}. 
We have done calculations for multiple values of $U$ and $\kappa$, and
our qualitative conclusions are similar for all cases. We
report the results for only $U/t = 3.33$ and $\kappa= 2$, since this
combination was found to be the most suitable for
PPV \cite{Chandross1}, and
it is likely that the Coulomb parameters in phenyl-based 
$\pi$-conjugated polymers and SWCNTs are similar. 

Full many-body calculation within Eq.~(1) is not possible for SWCNTs.
We use the single configuration interaction (SCI) approximation
\cite{Chandross,Rice,Chandross1}, 
which is a many-body approach valid within
the subspace of single excitations from the Hartree-Fock (H-F) ground
state.  While SCI is not sufficient for
two-photon states, semiquantitative results are obtained for
one-photon states.
We use open boundary condition along the nanotube (NT) axis, such that
evaluations of transition dipole matrix elements are simple.  
Surface states originating from ends of open tubes can be detected
from their energies at the chemical potential in the
$U\!\!=\!V_{ij}\!=\!0$ H\"uckel limit and their one-electron
wave functions, and they are excluded from the SCI calculations.
We have performed calculations for seven semiconducting zigzag ($n,0$)
NTs for $n$ ranging from 7 to 17, and (6,2), (6,4), and (7,6)
chiral NTs. 
The number of unit cells $N$ in SCI calculations for zigzag NTs is
18.  For the chiral NTs with large unit cells, we determined from
H\"uckel calculations the $N$ at which infinite system absorption
thresholds are reached, and performed the SCI calculations for these
$N$.  Our calculations are for $N=$10, 8, and 2 in the (6,2), (6,4),
and (7,6) NTs, with 1040, 1216, and 1016 C atoms, respectively.
%
%
\begin{figure}[ht]
  \includegraphics[width=1.8in,clip]{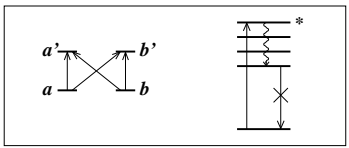}
\caption{ \label{fg1}
  (Left) Schematic of the four degenerate single-particle excitations
  from the highest occupied to the lowest unoccupied one-electron
  levels in SWCNTs.  (Right) These degeneracies are split by \Hee,
  and only the highest state is strongly dipole-allowed. Rapid relaxation
  occurs to the lowest forbidden exciton, radiative relaxation from which
  is forbidden.  
}
\end{figure}
%
%


We begin our discussions with the lowest energy excitations. 
In the zigzag SWCNTs 
the highest valence band (v.b.) and lowest conduction band (c.b.)  for
\Hee\ = 0 are doubly degenerate \cite{single-particle}.
In the chiral SWCNTs the degenerate
levels occur at different single-particle crystal momenta
\cite{single-particle}.
Nevertheless, in both cases there occur doubly degenerate
single-particle excitations with total crystal momentum zero.
Consider now the four degenerate lowest single-particle excitations in
SWCNTs, $\chi_{a \to a'}$, $\chi_{a \to b'}$, $\chi_{b \to a'}$, and
$\chi_{b \to b'}$, shown in Fig.~1, where $a, b~ (a', b')$ are the
highest occupied (lowest unoccupied) one-electron levels.
The two excitations $\chi_{a \to a'}$ and $\chi_{b \to b'}$ are
optically allowed, and for nonzero matrix elements of \Hee\ between
them, new nondegenerate eigenstates 
$\chi_{a \to a'} \pm \chi_{b \to b'}$ 
are obtained.  
There also occur superpositions involving the dipole forbidden
excitations,
as well as others involving immediately lower v.b.\ and higher c.b.\ 
levels.
Significantly, 
(i)  the odd superposition is dipole forbidden, and
(ii) for repulsive \Hee\ the allowed even superposition is higher in
energy, as is indicated in Fig.~1.
In Table~\ref{tb:E_mu} we have given the lowest SCI exciton state
energies and the squares of the transition dipole moments between them
and the H-F ground state, for the two representative cases of (11,0)
and (6,2) SWCNTs.
In both cases, the exciton state with strong dipole coupling is the
the highest energy excitation. In the chiral SWCNTs, there occur weakly
allowed states in between the strongly allowed state and the lowest
exciton state, but the overall behavior are similar.  
In Table~\ref{tb:all} we have listed ($n,m$) for all SWCNTs we have
investigated, and the corresponding differences in total energies
$\delta E$ between the optically allowed exciton and the lowest
exciton.

\begin{table}[ht]
\tabcolsep=14pt
\caption{                 \label{tb:E_mu}
  The energies of the lowest excitons and the squares of the transition 
  dipole couplings between them and the ground state $G$ 
  (electronic charge $e$=1). 
  The exciton at energy 1.259 eV in (11,0) is doubly degenerate. 
  Some of the forbidden excitons below the strongly allowed
  exciton in chiral SWCNTs  
  are odd superpositions of higher energy one-electron excitations.
}
\begin{ruledtabular}
\begin{tabular}{D{.}{.}{5}D{.}{.}{2}|D{.}{.}{4}D{.}{.}{2}}
%
\multicolumn{2}{c|}{(11,0)}    & \multicolumn{2}{c}{(6,2)} \\ 
%
\multicolumn{1}{c}{\rule[-4pt]{0pt}{14pt}$E_i$ (eV)} & 
\multicolumn{1}{c|}{$|\mu_{G,i}|^2$}                 &
\multicolumn{1}{c}{$E_i$ (eV)}                       & 
\multicolumn{1}{c}{$|\mu_{G,i}|^2$}                        \\ \hline
%
\rule{0pt}{9pt}
1.323      & 77.4 & 1.772 & 95.3   \\
1.321      &  0   & 1.768 &  0     \\
1.259\ (2) &  0   & 1.765 &  0     \\
1.231      &  0   & 1.764 & 13.5   \\
           &      & 1.743 &  0     \\
           &      & 1.743 &  0.327 \\
           &      & 1.733 &  0     \\
           &      & 1.710 &  0 
\end{tabular}
\end{ruledtabular}
\end{table}


\begin{table}[ht]
\caption{                 \label{tb:all}
Summary of computed SCI results for different SWCNTs.
}
\begin{ruledtabular}
\begin{tabular}{D{,}{,}{3}D{.}{.}{3}cccc}
\multicolumn{1}{c}{($n$,$m$)} & \multicolumn{1}{c}{$d$ (\AA)} & $\delta E$ (eV) & $E_{b1}$ (eV) & $E_{b2}$ (eV) & $E_{22}/E_{11}$ \\ \hline
 (7,0) &  5.56 & 0.113 & 0.540 & 0.782 & 1.801 \\
 (6,2) &  5.72 & 0.062 & 0.528 & 0.718 & 1.819 \\
 (8,0) &  6.35 & 0.098 & 0.533 & 0.578 & 1.646 \\ 
 (6,4) &  6.92 & 0.057 & 0.480 & 0.552 & 1.716 \\
(10,0) &  7.94 & 0.126 & 0.406 & 0.574 & 1.650 \\
(11,0) &  8.73 & 0.092 & 0.415 & 0.454 & 1.726 \\
 (7,6) &  8.95 & 0.073 & 0.365 & 0.470 & 1.675 \\
(13,0) & 10.3  & 0.113 & 0.322 & 0.454 & 1.577 \\
(14,0) & 11.1  & 0.089 & 0.338 & 0.386 & 1.677 \\
(17,0) & 13.5  & 0.086 & 0.288 & 0.312 & 1.698
\end{tabular}
\end{ruledtabular}
\end{table}

The energy spectra of SWCNTs is similar to that of polyacetylenes and
polydiacetylenes, where also there occur dipole-forbidden excited
states below the optical exciton as a consequence of e-e interaction
\cite{Hudson}.  
PL is weak in these polymers, as the optically excited
state decays in ultrafast times to the low energy dipole-forbidden
state, radiative transition from which to the ground state cannot
occur. 
The results of Tables~\ref{tb:E_mu} and \ref{tb:all} then strongly
suggest that the low QE of PL in SWCNTs ($< 10^{-3}$)
\cite{OConnell-02-science,Lebedkin03,Wang04,Sheng04} is intrinsic.
(The one-photon forbidden state in the polymers is two-photon allowed,
while the lower energy states in the SWCNTs are not. This difference
is of no consequence in emission, which is a one-photon process.)  
We will return to this issue later.


%
\begin{figure}[ht]
  \includegraphics[width=2.5in,clip]{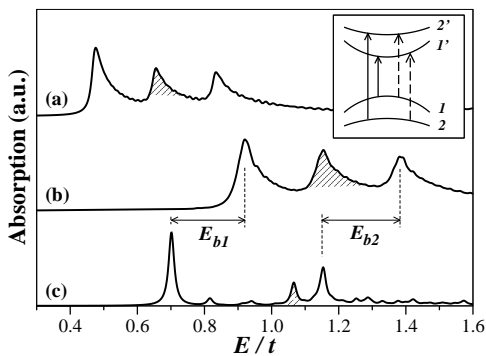}
\caption{ \label{fg:NT80}
  Absorption spectra of (8,0) nanotube from (a) H\"uckel, (b)
  Hartree-Fock, and (c) SCI calculations. Shaded peaks indicate
  transverse polarized absorptions.
  The inset shows 
  longitudinal (solid arrows) and transverse (dashed arrows)
  excitations.
}
\end{figure}
%
%

Within \TB\ theory, excitations responsible for optical absorptions
polarized transverse to the tube axis, 
$\chi_{1 \to 2'}$ and $\chi_{2 \to 1'}$ (see inset, Fig.~2), 
are also degenerate.  
The threshold energy for transverse excitation is exactly halfway
between the energies of the two longitudinal excitations 
$\chi_{1 \to 1'}$ and $\chi_{2 \to 2'}$.  
\Hee\ will also split the degeneracy among the transverse excitations
(as mentioned above, we ignore here the depolarization effect
\cite{AjikiAndo94}, as the splitting due to many-body effects will
occur independent of the intensity of transverse absorptions).
We now expect a dipole-forbidden transition 
$\chi_{1 \to 2'} - \chi_{2 \to 1'}$ 
shifted below the central region and a dipole-allowed transition
$\chi_{1 \to 2'} + \chi_{2 \to 1'}$ 
shifted above the central region.  
In Fig.~2 we have shown the calculated optical absorptions within \TB,
H-F, and SCI approaches for the (8,0) SWCNT.  
Strong blueshift of the dipole-allowed transverse excitation from the
central region is seen.  
Very similar relative blueshift of the transverse optical excitation
has been of strong theoretical \cite{Chandross,Rice,Chandross1} and
experimental \cite{Chandross,Comoretto} interest in PPV.  
Detection of this blueshift in SWCNTs can give a measure of the
strength of the \ee\ interaction.


We now focus on the observable parallel absorptions.  
Our calculated spectra in all cases resemble the three spectra in
Fig.~2 for the (8,0) NT.  
Within SCI the H-F thresholds are the edges of the continuum bands
corresponding to each class of excitations.  Thus in Fig.~2 $E_{b1}$
and $E_{b2}$ are the binding energies of the two lowest excitons.  We
examine $E_{22}/E_{11}$, where $E_{11}$ and $E_{22}$ are the energies
of the two lowest longitudinal absorptions.  
The \TB\ $E_{22}/E_{11}$ is close to 2 for the (8,0) NT.  
Figure~2 indicates that even at the H-F level $E_{22}/E_{11} < 2$. 
The simple reason is that unless the correlation-induced blueshift of
$E_{22}$ is twice that of the $E_{11}$, the ratio is bound to be less
than 2.  
As seen in Fig.~2 the energy shifts are nearly the same for both
absorption features of the (8,0) NT, at both H-F and SCI levels.  
We have found this to be true for all SWCNTs that we have
investigated.  
As shown in Table~\ref{tb:all} the SCI $E_{22}/E_{11}$ for large
diameter NTs is close to the experimental value of $\sim$ 1.7
\cite{OConnell-02-science}.

Each optical exciton in the SWCNTs has its own binding energy, as
shown in Fig.~\ref{fg:NT80} for the (8,0) NT. 
The occurrence of high energy bound excitons beyond the continuum
threshold corresponding to the lowest excitation is also known in
$\pi$-conjugated polymers with multiple one-electron bands
\cite{Chandross,Rice}.  
In Table~\ref{tb:all} we have listed the exciton binding energies
$E_{b1}$ and $E_{b2}$ for different SWCNTs.  For the narrowest NTs,
$E_{b2} > E_{b1}$, while for the widest NTs $E_{b2} \simeq E_{b1}$.
We have done calculations for four different $U$ ($U/t$=1.9, 2.5,
3.33, and 4.0), and two values of $\kappa$ ($\kappa$=1 and 2).
The general features of 
(i)   decreasing $E_{b1}$ and $E_{b2}$ with increasing diameter $d$, 
(ii)  $E_{b2} > E_{b1}$ for the narrowest NTs, and 
(iii) $E_{b2} \simeq E_{b1}$ in the widest NTs 
are true for all parameters.

In Ref.~\onlinecite{Chandross1} we had shown that the combination $U$
= 8.0 eV and $\kappa$ = 2 (out of a total of fifteen sets) gave the
best fits to four different absorption bands in PPV, and that with
this parameter set the calculated exciton binding energy is $\sim$ 0.9
$\pm$ 0.2 eV. Very similar magnitude was subsequently calculated
within an \textit{ab initio} approach that included solution of the
Bethe-Salpeter equation for the two-particle Green's function
\cite{Rohlfing99}.  Using the same $U$ and $\kappa$ we find that the
$E_{b1}$ in the widest SWCNTs in Table~\ref{tb:all} are about 0.3 eV,
while in the narrower NTs $E_{b1} \sim 0.5$ eV. Indeed, for all eight
combinations of $U$ and $\kappa$ we found that the exciton binding
energies in the SWCNTs are smaller than in PPV.
%
%
Conwell has suggested that the usual definition of the exciton binding
energy may not be appropriate for PPV, and that the exciton binding
energy should be defined as the energy required to dissociate the
exciton into a separated pair of oppositely charged polarons
\cite{Conwell}. It has been claimed that taking into account the
relaxation energy of the polarons gives an exciton binding energy as
small as 0.4 eV in PPV \cite{Conwell}. Further work is required to
determine whether such a correction would be appropriate also for
SWCNTs.

%
%
\begin{figure}[ht]
  \includegraphics[width=2.4in,clip]{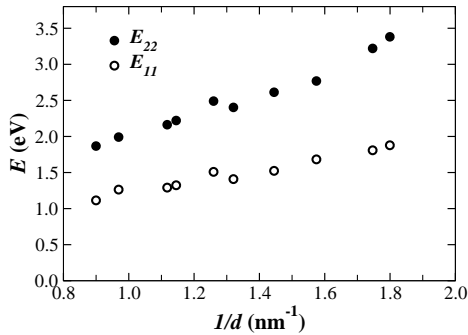}
\caption{ \label{fg:peak_vs_d}
  Calculated energies of the two lowest excitons,
  versus inverse diameter of the SWCNTs.
}
\end{figure}
%
%
In Fig.~\ref{fg:peak_vs_d} we have plotted the calculated SCI $E_{11}$
and $E_{22}$ for all ten SWCNTs against their inverse diameters $1/d$.
The linear decrease of the optical excitation energies with decreasing
$1/d$, observed experimentally, is well reproduced. The absolute
energies are larger than what are experimentally observed, as
expected, as SCI is not entirely a quantitative method.  
This can also indicate that e-e interactions in the SWCNTs are
somewhat smaller.


We now return to our conclusion that the low QE of PL ($<10^{-3}$)
\cite{OConnell-02-science,Lebedkin03,Wang04,Sheng04} in SWCNTs is an
intrinsic feature of isolated NTs, and not a consequence of exciton
quenching.
As shown in Table~\ref{tb:all}, $\delta E$ $\sim$ 3--4 $k_B T$.  Thus
following the rapid relaxation into the forbidden lowest exciton it is
unlikely that thermal effects will re-excite the system to the allowed
state.  Intrinsic radiative decay rate therefore should be low, and
the radiative lifetimes large.
Simultaneously,
$\delta E$ is small enough that small amounts of impurities or
changes in the environment can modify the emissive behavior.
This may explain the strong dependence of the emission on the
environment \cite{OConnell-02-science,Lebedkin03,Wang04,Lefebvre03}.
Recent estimates of very long exciton lifetimes
\cite{Wang04,Sheng04} are in agreement with our work.
Femtosecond time-resolved measurements indicate same decay rates for
fluorescence and PB, but the PB drops to only half its peak value
\cite{Wang04}. 
We agree with Ref.~\onlinecite{Wang04} that this is an indication of
trapping of the excitation in a non-emissive state. We also believe
that the non-emissive state is the forbidden exciton found here.

In summary, semiempirical configuration interaction calculations
reveal excitonic electronic structures for SWCNTs, and also allows
direct comparisons to $\pi$-conjugated polymers.  Corresponding to
each band-to-band transition within \TB\ theory there occurs an
optical exciton in SWCNTs. The ratio problem is a simple consequence
of nearly equal blueshifts of the two lowest optical absorptions from
their \TB\ frequencies. The binding energies of the lowest two
excitons decrease with increasing diameter and the two binding
energies are comparable for wide NTs. Assumption of similar Coulomb
parameters in SWCNTs and phenyl-based $\pi$-conjugated polymers gives
smaller binding energy for the former.  We estimate 0.3--0.5 eV
binding energy for the wide SWCNTs. We ascribe the low QE of the PL in
SWCNTs to the occurrence of optically forbidden excitons below the
optical exciton, which in turn is a consequence of the splitting of
the degeneracy that exists in the one-electron limit by e-e
interactions. A similar degeneracy splitting should also occur between
the states to which optical excitations transverse to the NT axis
occurs.

This work was supported by NSF-DMR-0406604, NSF-ECS-0108696 and the
NSF STC at the University of Arizona. We acknowledge useful
discussions with Professors Yu.~N. Gartstein, E.~J. Mele, and Z.~V.
Vardeny.

\textit{Note added.} The occurrence of forbidden excitons below the lowest
optical exciton in zigzag SWCNTs, and the similarity in the binding
energies of the first two optical excitons have been found in a 
recent Letter \cite{Perebeinos04}. Recent work has also claimed that the
dominant source of the blueshift of the optical absorption and the 
ratio problem is the Coulomb self energy \cite{Kane04}, in agreement with 
the work presented here.


\end{document}